\documentclass[twocolumn,aps,showpacs,floatfix,prc]{revtex4}
\usepackage[dvips]{epsfig}
\begin{document}

\title{Comparison of Yields of neutron rich nuclei in Proton and Photon induced $^{238}$U fission}
\author{ F. A. Khan$^{1**}$, Debasis Bhowmick$^{2*}$,  D. N. Basu$^{3*}$, M. Farooq$^{4**}$ and Alok Chakrabarti$^{5*}$}

\address{ $^*$ Variable  Energy  Cyclotron  Centre, 1/AF Bidhan Nagar, Kolkata 700 064, India}
\address{ $^{**}$ Dept. of Physics, University of Kashmir, Hazratbal, Srinagar 190 006, India}

\email[E-mail 1: ]{29firdous11@gmail.com}
\email[E-mail 2: ]{dbhowmick@vecc.gov.in}
\email[E-mail 3: ]{dnb@vecc.gov.in}
\email[E-mail 4: ]{farooq@gmail.com}
\email[E-mail 5: ]{alok@vecc.gov.in}
\date{\today}

\begin{abstract}

    	A comparative study of fission of actinides specially $^{238}$U, by proton and bremsstrahlung photon is performed. Relative mass distribution of $^{238}$U fission fragments have been explored theoretically for both proton and photon induced fission. The integrated yield along with charge distribution of the products are calculated to find out the neutron richness in comparison to the nuclei produced by r-process in nucleosynthesis. Some r-process nuclei in intermediate mass range for symmetric fission mode are found to be produced almost two order of magnitude more for proton induced fission than photofission, although rest of the neutron rich nuclei in the asymmetric mode are produced in comparable proportion for both the processes.  
\vskip 0.2cm
\noindent
{\it Keywords}: Photonuclear reactions; Photofission; Nuclear fissility; GDR; Exotic nuclei.   
\end{abstract}

\pacs{ 25.20.-x, 27.90.+b, 25.85.Jg, 25.20.Dc, 29.25.Rm }  
 
\maketitle

\noindent
\section{Introduction}
\label{section1}

    Fission of actinide targets, especially Uranium or thorium is a highly promising route for producing, neutron rich (n-rich) Radioactive Ion Beam (RIB) for nuclear spectrometry. Fission by neutron \cite{Na13} , proton \cite{Depp13} and photon \cite{Is14,Be15} has been studied for several decades and is still relevant. Among three, photo-fission \cite{Es03} has been found more promising for having better thermal management and being a cold process creates higher yields for n-rich nuclei as compared to light ion induced fission except in the mass range of $110<A<125$. Therefore there is a renewed interest presently to go for photo-fission \cite{Di99} using e-LINAC as a primary accelerator to produce energetic photons in the Giant Dipole Resonance (GDR) region \cite{Bh15, Bh16}. The Advanced Rare IsotopE Laboratory (ARIEL) \cite{TRIUMF12,TRIUMF11} at TRIUMF, JINR, Dubna \cite{Dubna} and at ALTO, IPN, Orsay \cite{Alto} are the laboratories, where initiative have already been taken. As an extension of the present RIB development, a facility called ANURIB \cite{Ch13} (Advanced National facility for Unstable and Rare Isotope Beams) will be coming up at this center with e-LINAC as primary accelerator for photo-fission. However, it is true that low energy proton induced fission also provides relatively less expensive means to produce n-rich nuclei substantially in a specific mass region. For this one can have low energy proton beam (either from cyclotron or proton LINAC) instead of e-LINAC, so that n-rich RIBs, produced in 2nd target station be put subsequently, in the same post-accelerator module.

	It is well known that the properties of fission mass distribution is governed by asymmetric and symmetric mass split of the fissioning nuclei depending on the excitation energy. In the energy range 12-30 MeV, $^{238}$U fission is known to take place in both asymmetric and symmetric fission modes with comparable probabilities \cite{Jones}. However probability of asymmetric fission relative to that of symmetric fission decreases with proton energies. For very high energy (100 MeV), the mass yield curve does not have any trough, although the total fission cross section remains constant after 30 MeV \cite{Stev}. In neutron and photon induced fission, at low energies, the pattern of mass distribution of the products are found to be nearly identical with that of proton induced fission. However, in case of photo-fission, the n-rich nuclei produced in symmetric fission mode have much lower cross section compared to two asymmetric modes and increasing the excitation energy does not help because total cross-section decreases rapidly due to absence of giant dipole resonance, unlike proton induced fission where increase of symmetric fission mode has been compensated by reduction in asymmetric mode, so that total fission cross section remains unchanged after 30 MeV. Therefore, in the present work, we would like to perform a simultaneous analysis of the behavior of the symmetric and asymmetric modes of proton induced fission for different excitation energies of $^{238}$U and a comparison with that of photo-fission is made. Depending on the availability of data, the analysis can also be be extended to other actinides (Th, Pu, Am, Np etc.). TALYS \cite{Talys} and PACE4 \cite{Pace} results are also incorporated in the comparison for the calculation of total cross section of proton induced fission along with experimental data because both the codes take into account the competition of other reaction channels in addition to fission. Finally the role of proton induced fission of actinides, specially $^{238}$U towards the production of n-rich nuclei in the mass range 110$<A<$125 is explored. 
      
\noindent
\section{Theoretical Formalism of Fission Cross-section}
\label{section2}    
  
    The empirical formula employed in our calculation is taken from Ref.\cite{Proko} as  
    
\vspace{-0.1cm}
\begin{equation}
\sigma_f (E_p)= P_1[1 - \exp(-P_2(E_p - P_3))]
\label{seqn1}
\vspace{-0.1cm}
\end{equation}
\noindent
where $\sigma_f$, is the total fission cross-section (mb); $E_p$ is the incident proton energy (MeV). $P_i (i=1,2,3)$ are the arbitrary fitting parameters with physical meaning that $P_1$,the saturation cross-section, $P_2$, the increasing rate of cross-section with energy and $P_3$, the apparent threshold energy respectively. The $P_i$'s have been parametrized with fissility parameter $Z^2/A$ as:

\vspace{-0.1cm}
\begin{equation}
P_i (Z^2/A)= \exp[Q_{i,1} + Q_{i,2}( Z^2/A) + Q_{i,3}( Z^2/A)^2]
\label{seqn2}
\vspace{-0.1cm}
\end{equation}
\noindent
where $Q_{i}$'s are the coefficients of the powers of $(Z^2/A)$ which are also determined by fitting the experimental fission cross-section data for a wide range of fissioning nucleus from $^{181}Ta$ to $^{181}Bi$ as shown by Fukahori and Pearlstein \cite{Fuka91}. However, later on Fukahori and Chiba \cite{Fuka97} modified the expression, where the fission probability was calculated as a ratio of experimental fission cross-section and the total reaction cross-section as calculated by Letaw \cite{Letaw}. With the availability of precise data of fission cross-sections from time to time, the systematics improved quite a lot. Systematics used by Prokofiev \cite{Proko} is found to be very effective in the energy range 12-63 MeV for calculation of total fission cross-section $\sigma_f$ as described in Eq.(2) by fitting experimental data \cite{Baba, Isa08}, while $P_1$ is parametrized differently as:

\vspace{-0.1cm}
\begin{equation}
P_1 (Z^2/A)= {R_{11}} {[1 - {\exp(-{R_{13}}(Z^2/A  - {R_{12}}))}]}
\label{seqn3}
\vspace{-0.1cm}
\end{equation}
\noindent

It is important to mention here that we have not considered here the high energy correction term because it is effective for proton energy in the range hundreds of MeV or more. The values of $P_2 = 0.111$ and $P_3 = 12.1$ are considered as constant since from Ref.\cite{Fuka91}, it is found to be invariant for $ Z^2/A$ in the range 35.9 to 36.1 for $^{132}$Th - $^{239}$Pu. The values of $R_1j(j=1-3)$are fitted by least square fit and found to be $R_{11} = 2730 \pm 82.962$, $R_{12} = 34.99 \pm 0.034$ and $R_{13} = 2.07 \pm 0.120$.
         
\noindent
\section{Systematics of Mass and Charge Distribution}
\label{section3}

    Although we have tried to accumulate the data till date as much as possible \cite{Oht,Rub}, the availability of data for both fission cross-section and mass distribution are not too many in the energy range of 13-60 MeV. In general the mass distribution is interpreted as a sum of the contribution from the symmetric and asymmetric fission modes for multimode fission model. Each fission mode corresponds to the passage through the fission barrier of specific shape. For each fission mode, the yield is described in the form of a Gaussian function. Three Gaussian functions are found to be good enough for describing three fission modes. The symmetric fission mode (SM) is peaked around A = 118, while for asymmetric fission modes (ASYM), the maxima are at A = 137 and A = 99. The total yield of fragments whose mass number is A is given by the expression:
    
\begin{eqnarray}
 Y(A) =&& Y_{SM}(A) + Y^1_{ASYM}(A) + Y^2_{ASYM}(A) \nonumber\\
 =&& C_{SM}\exp\Big[-\frac{(A-A_{SM})^2}{2\sigma^2_{SM}} \Big] \nonumber\\
 && + C_{ASYM}\exp\Big[-\frac{(A-A_{SM} - D_{ASYM})^2}{2\sigma^2_{ASYM}} \Big] \nonumber\\
&& + C_{ASYM}\exp\Big[-\frac{(A-A_{SM} + D_{ASYM})^2}{2\sigma^2_{ASYM}} \Big] \nonumber\\ 
\label{seqn4}
\end{eqnarray}
\noindent
where, the Gaussian function parameters, $C_{SM}$, $C_{ASYM}$ and $\sigma _{SM}$, $\sigma _{ASYM}$ are the amplitudes and widths, respectively, of the symmetric (SM) and asymmetric (ASYM) fission modes, and $A_{SM}$ is the most probable mass value for the symmetric fission mode while $A_{SM} - D_{ASYM}$ and $A_{SM} + D_{ASYM}$ being the most probable masses of a light and the complementary heavy fragments respectively in the asymmetric fission mode. 

\begin{figure}
\vspace{0.0cm}
\eject\centerline{\epsfig{file=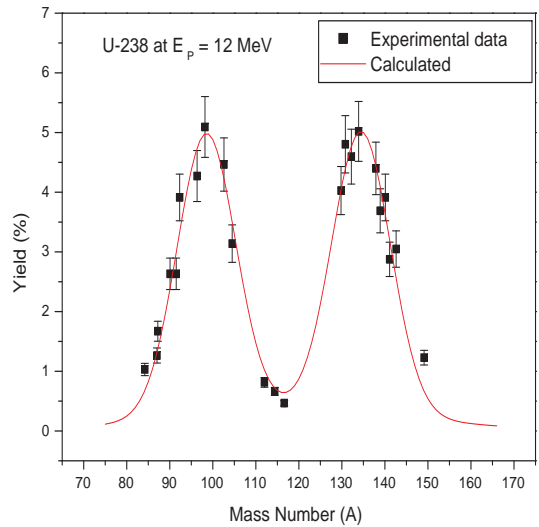,height=7cm,width=7cm}}
\caption
{Comparison of the measured mass yield distribution (full circles) \cite{Oht} for $^{238}$U fission induced by 12 MeV Proton with the prediction (solid line) of the three Gaussian formula for $Y(A)$.}
\label{fig1}
\vspace{0.5cm}
\end{figure}
\noindent

\begin{figure}[htbp]
\vspace{0.0cm}
\eject\centerline{\epsfig{file=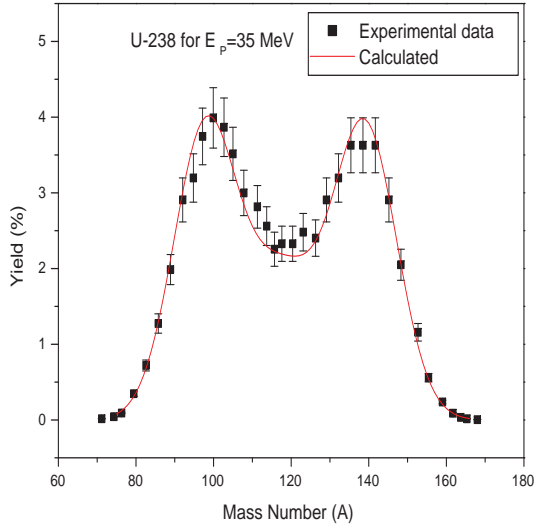,height=7cm,width=7cm}}
\caption
{Same as Fig.1(full circles) \cite{Rub} but for different proton energy of 35 MeV.} 
\label{fig2}
\vspace{0.5cm}
\end{figure}
\noindent

	In Figs.1-2, approximation by the preceding three Gaussian functions for the mass distribution Y(A) of fragments per 100 fission events originating from $^{238}$U fission induced by 12  and 35 MeV proton are plotted and compared with experimental data in Ref.\cite{Oht} and Ref.\cite{Rub} respectively. It is important to mention here that with increasing energy, the amplitude of both the asymmetric peak decreases, although not so appreciably in the energy domain considered here. However, as far as symmetric fission is concerned, the increase is substantial. Therefore, the probability of symmetric fission relative to that of symmetric fission decreases appreciably with increasing proton energy. The values of $A_{SM}$, $D_{ASYM}$ are 118, 19 respectively, whereas the variation of four other parameters with proton energy ranging from 12 to 50 MeV are plotted in Fig.3. The lines represent least square fits assuming quadratic energy dependence of the parameters. It is important to note that the value of $\sigma_{SM}$ at 12 MeV is not shown in Fig.3 because of large error due to insufficient experimental data points. Furthermore, the percentage yield of isotopes having mass number in the range 115-120 are very small and due to which the fitted mean value of $\sigma_{SM}$ is found to be very high ( e.g. $\sigma_{SM}$ = 27.7 $\pm$ 14.2 for Fig.1). In fact, at such a low energy, fission of $^{238}$U is perfectly asymmetric and one can fit with two Gaussian function leaving aside the symmetric term in the expression of mass yield distribution Y(A) in Eq.(4).
	 
\begin{figure}
\vspace{0.0cm}
\eject\centerline{\epsfig{file=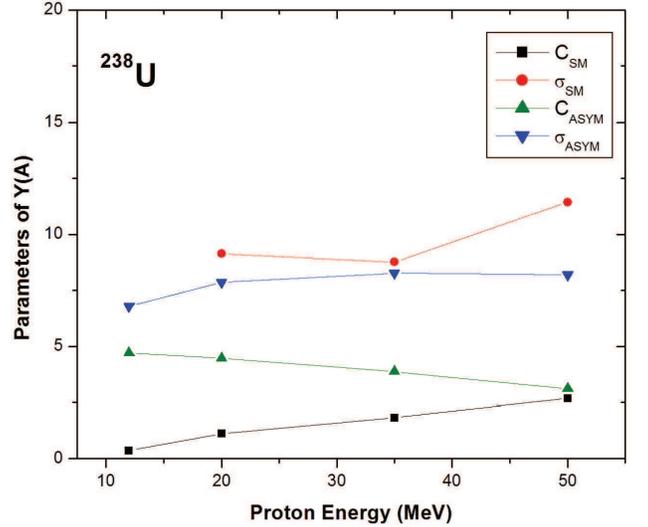,height=7cm,width=7cm}}
\caption
{Variation of symmetric and asymmetric Gaussian parameters with excitation energies.}
\label{fig3}
\vspace{0.5cm}
\end{figure}
\noindent

\begin{figure}[htbp]
\vspace{0.0cm}
\eject\centerline{\epsfig{file=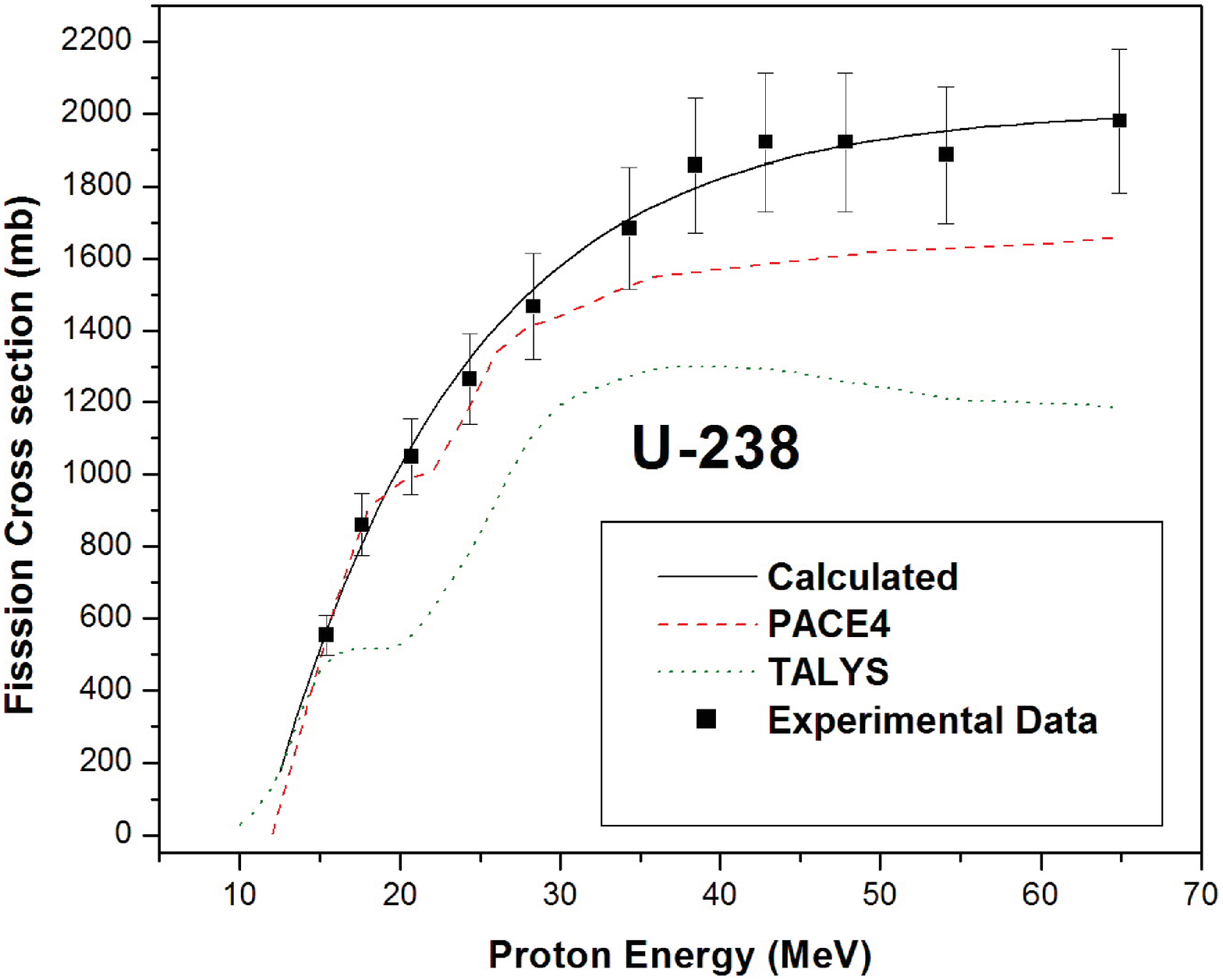,height=7cm,width=7cm}}
\caption
{Comparison of total fission cross-section (full circles) \cite{Baba, Isa08} of $^{238}$U by proton with energies in the range 13-63 MeV with the prediction (solid line) from our calculation, TALYS and PACE4.} 
\label{fig4}
\vspace{0.5cm}
\end{figure}
\noindent

    The isobaric charge distribution of photofission products can be well simulated by a single Gaussian function as:

\begin{equation}
 Y(A,Z) = \frac{Y(A)}{\sqrt{\pi C_p}} \exp\Big[-\frac{(Z-Z_s)^2}{C_p} \Big]
\label{seqn5}
\end{equation}
\noindent    
where, $Z_s$ represents most stable isotope of fission fragment with mass number $A$. In order to deduce expression for $Z_s$, theoretically, for the most stable nucleus by keeping mass number $A$ constant while differentiating liquid drop model mass formula and setting the term $\partial M_{nucleus}(A,Z)/\partial Z\mid_A$ equal to zero as described in Ref.\cite{Ch06}. The value of the parameter $C_p$ which decides the dispersion for the most probable isotope is extracted by fitting experimental data \cite{Baba} is found to be 0.95. In the present work the form of the charge distribution is assumed to be independent of the mass number of the actinides and proton energy for the range considered here. The atomic numbers $Z_s$ used in Eq. (5) for the most stable nuclei are calculated using values $a_c$ = 0.71 MeV and $a_{asym}$ = 23.21 MeV \cite{Ch06}. 

\begin{figure}[htbp]
\vspace{0.0cm}
\eject\centerline{\epsfig{file=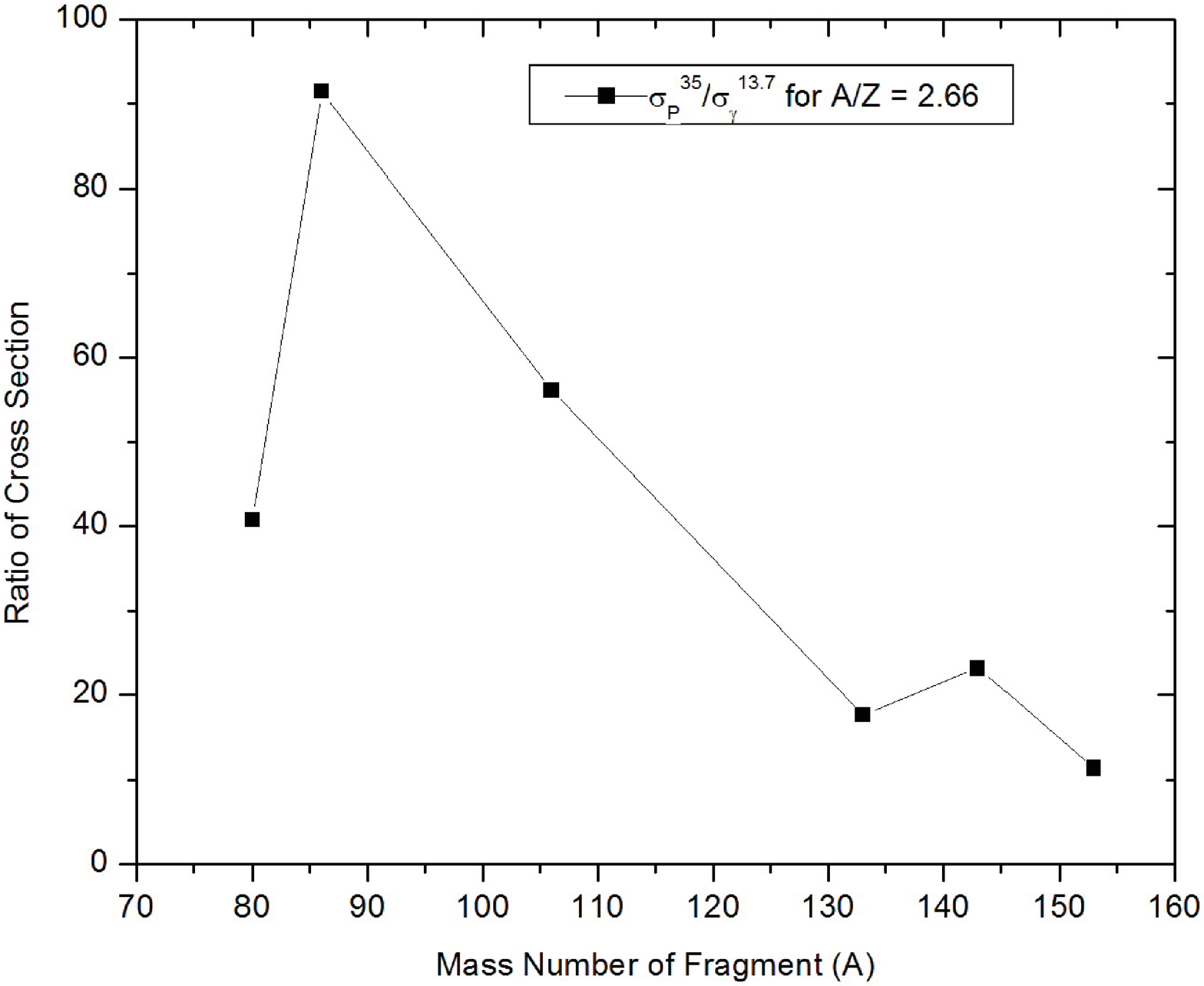,height=7cm,width=7cm}}
\caption
{Variation of ratio of cross sections (35 MeV p/13.7 MeV photon) with fragment mass number for A/Z = 2.5.} 
\label{fig5}
\vspace{0.5cm}
\end{figure}
\noindent

\noindent
\section{Calculation and Results}
\label{section4}

		The production cross sections of individual fission fragments induced by protons are obtained by multiplying fission cross section, $\sigma_f(E_p)$, as calculated by the empirical formula in Eq.(1), by charge distribution which means: $\sigma_f(A,Z)= \sigma_f(E_p).Y(A,Z)/100$. 
		
		It is important to mention here that total fission cross-section $\sigma_f(E_p)$ increases with increasing proton energy, but it saturates almost at 30 MeV. However, for photofission, the cross section maximizes in the GDR range for mean photon energy of 13.7 MeV. In Fig.4 $\sigma_f(E_p)$ vs  proton energy has been plotted using PACE4 \cite{Pace} and TALYS \cite{Talys} along with our empirical formalism to compare with experimental data \cite{Baba, Isa08}. In an attempt to make a comparison of proton induced fission with photo-fission, the choice of energy of the incident particle (proton/photon) is made such that the fission cross-section is maximized. 

\begin{figure}
\vspace{0.0cm}
\eject\centerline{\epsfig{file=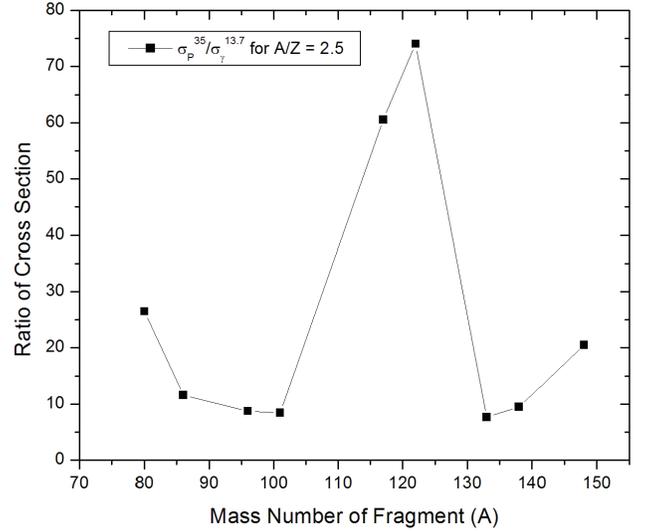,height=7cm,width=7cm}}
\caption
{Same as Fig.4 but for A/Z = 2.66.} 
\label{fig6}
\vspace{0.5cm}
\end{figure}
\noindent
	
		In Figs.5-6, the ratio of cross-sections of proton and photon induced fission at energies 35 MeV and 13.7 MeV respectively as $(\sigma_p^{35}/\sigma_{\gamma}^{13.7})$ for two different A/Z at 2.50 and 2.66 are plotted for a range of fragment mass number from 80 to 150. It is evident from Fig.5 that in general proton induced fission cross-section is an order magnitude higher than photo-fission in the asymmetric mode, while in symmetric fission mode the order of enhancement is more than 70-90 times. Moreover, as A/Z of fissioning nucleus increases the peak value of the ratio shifts towards lower mass number in the symmetric fission mode. For A/Z = 2.50 the ratio of the cross-section peaks around A (product) = 120, while for A/Z = 2.66 it is around 96 (Figs.5-6). 

\begin{figure}[htbp]
\vspace{0.0cm}
\eject\centerline{\epsfig{file=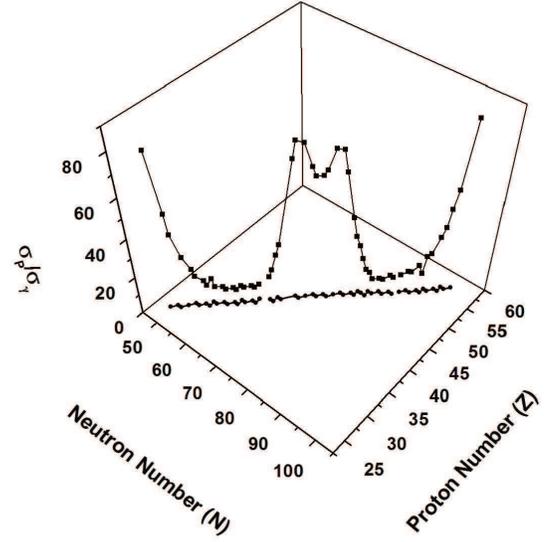,height=7.2cm,width=6.59cm}}
\caption
{Three dimensional plot for the ratio of proton and $\gamma$ induced $^{238}$U fission cross sections vs atomic number Z and neutron number N. The two dimensional projection on the N-Z plane is also shown.}
\label{fig7}
\vspace{0.5cm}
\end{figure}
\noindent	
  
\begin{table*}[htbp] 
\vspace{0.0cm}
\caption{\label{tab:table1} The theoretical cross sections for production of nuclei with same A/Z.}
\begin{tabular}{|c|c|c|c|c|c|c|c|c|}
\hline
Nuclei& $A/Z$& $Z_s$ & $A_s$ &$Z_s$-Z& $A^F$-$A^F_s$& ${\sigma}_\gamma^{13.7}$ (mb)& $ {\sigma}_P^{35}$(mb)& $ {\sigma}_P^{35}$ /${\sigma}_\gamma^{13.7}$ \\
\hline

$^{80}$Zn &   2.66&   36&   64  & 6&   16&   $2.6\times 10^{-3}$& $1.1\times 10^{-1}$&40.00  \\ \hline

$^{96}$Kr &   2.66&   42&   84  & 6&   12&  $2.1\times 10^{-3}$& $1.9\times 10^{-1}$&91.43  \\ \hline

$^{106}$Zr&   2.66&   46&   91  & 6&   15&  $3.9\times 10^{-3}$& $2.2\times 10^{-1}$& 56.15  \\ \hline

$^{133}$Sn&   2.66&   56&   119 & 6&   14&  $2.2\times 10^{-3}$& $3.9\times 10^{-2}$& 17.59  \\ \hline

$^{143}$Xe&   2.66&   60&   131 & 6&   12&  $6.6\times 10^{-3}$& $1.5\times 10^{-1}$&  23.18  \\\hline

$^{154}$Ce&   2.66&   64&   140 & 6&   14&  $2.2\times 10^{-3}$& $2.5\times 10^{-2}$& 11.32 \\ \hline

\hline
\end{tabular} 
\vspace{0.0cm}
\end{table*}  
 
\begin{table*}[htbp]
\vspace{0.0cm}
\caption{\label{tab:table2} The theoretical cross sections for production of some r-process nuclei.}
\begin{tabular}{|c|c|c|c|c|c|c|c|c|}
\hline
Nuclei&$A/Z$&$Z_s$ &$A_s$ &$Z_s$-$Z$&$A^F$-$A^F_s$&${\sigma}_\gamma^{13.7}$ (mb)& $ {\sigma}_P^{35}$(mb)&$ {\sigma}_P^{35}$ /${\sigma}_\gamma^{13.7}$ \\
\hline

$^{80}$Ge &   2.50&   36&   74  & 4&   6&   $1.2\times 10^{-1}$& 3.07 & 26.53  \\ \hline

$^{86}$Se &   2.52&   38&   80  & 4&   6&  $1.1\times 10^{0}$& 12.42& 11.62  \\ \hline

$^{96}$Sr &   2.53&   42&   88  & 4&   8&  $4.4\times 10^{0}$& 38.38& 8.77  \\ \hline

$^{101}$Zr&  2.53&   44&   90 & 4&   11&  $4.8\times 10^{0}$& 40.74& 8.46  \\ \hline

$^{117}$Pd&   2.54&  50& 106 & 4&   9&   $3.3\times 10^{-1}$& 19.86& 60.56  \\\hline

$^{122}$Cd&   2.54& 114&  84 & 4&   8&  $2.9\times 10^{-1}$& 21.52 & 74.04 \\ \hline

$^{133}$Te&   2.56&   56& 128 & 4&  5&   $3.7\times 10^{0}$& 28.55 & 7.68 \\ \hline

$^{138}$Xe&   2.55&   58&  132& 4&  6&   $3.9\times 10^{0}$& 37.89 & 9.53 \\ \hline

$^{148}$Ce&   2.55&   62&  138& 4& 10&  $1.1\times 10^{0}$& 22.77& 20.50 \\ \hline

\hline
\end{tabular} 
\vspace{0.0cm}
\end{table*}
\vspace{0.0cm} 

\begin{figure}[htbp]
\vspace{0.4cm}
\eject\centerline{\epsfig{file=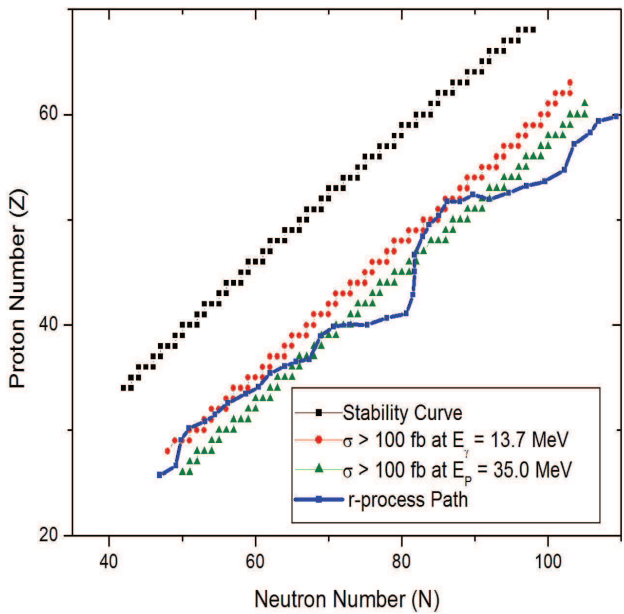,height=8.1cm,width=7cm}}
\caption
{Plots of atomic number Z vs neutron number N for exotic nuclei produced by photon and proton induced fission of $^{238}$U.}
\label{fig8}
\vspace{0.0cm}
\end{figure}
\noindent	

    In Fig.7, comparison between proton induced fission and photo-fission has been carried out in a different way where the ratio of proton and $\gamma$ induced $^{238}$U fission cross section are plotted in neutron  number (N) and atomic number (Z) plane. The cross sections for most neutron rich isobars are calculated  with proton and mean photon energy of 35 MeV and 13.7 MeV respectively subject to a limit of cross section  $>$100fb. The two dimensional projection on the N-Z plane is also shown. It is evident from Fig.7 that  the enhancement of the ratio of cross section $(\sigma_p^{35}/\sigma_{\gamma}^{13.7})$ is substantial  of the order $\sim$ 80 in the mass range A = 110$-$125 along with two fringes A = 75 and A = 160, while it is $\sim$ 10 for the remaining masses. The experimentally observed $\beta$ stable nuclei along with the r-process nuclei are also shown in Fig.8 in order to highlight how much one can march away from $\beta$ stability towards r-process path using $^{238}$U proton and photon induced fission. Although, there is no significant difference between proton and photon induced fission so far production of neutron rich nuclei are concerned, still proton induced fission is little ahead towards n-drip line than other one.  
	
		In an effort to investigate the production cross-section of neutron-rich nuclei by proton and photon induced fission, some isotopes with maximum cross-section (appearing at the two asymmetric peaks of the mass distributions), some with small lower cross-section (appearing at the symmetric mass distribution) are arranged in Table 1 with proton and average photon energies 35 and 13.7 MeV respectively. It is evident from Table1, that relative enhancement proton induced fission over photon is substantially high for products in the symmetric mass distribution domain (e.g. for $^{122}$Cd). Moreover, in Table 2 the production cross-sections of some r-process nuclei are highlighted where cross-sections are tabulated both for proton and photon induced fission. For some waiting point nuclei: e.g., $^{80}$Zn and $^{134}$Sn, the production cross-sections are enhanced by 40 and 20 times respectively by proton than to photon induced fission.   Comparing Tables 1 and 2, one may notice that for A/Z almost equal to 2.55 $\pm$ 0.01, when $Z_s - Z$ = 4, the cross sections are few tens of mili-barns, while for A/Z almost equal to 2.66, when $Z_s - Z$ = 6 the cross sections reduces to one tenth of micro-barn for proton induced fission of $^{238}$U. The reduction of cross section is consistent because cross sections fall rapidly with increasing mass number because of the neutron richness that is obvious from Eq.(5).
		
\noindent
\section{Summary and conclusion}
\label{section4}

	In summary, we find that the two mode fission mechanism with three Gaussian function along with four arbitrary parameters behave quite satisfactorily on the observed mass yield curve for proton induced fission of $^{238}$U up to 60 MeV. Empirical formalism nicely reproduces total fission cross-section for fission of various actinide elements in a wide range of incident proton energy and in the high energy domain specially, it fits better than TALYS and PACE4 codes, where competition of other reaction channels are considered in addition to fission.
	
	As far as the production of neutron rich nuclei is concerned, comparisons have been carried out in between proton and photon induced fission of $^{238}$U. The present calculation indicates clearly that many of the r-process nuclei in intermediate mass range can be obtained in the laboratory with measurable cross-section both by proton and photon induced fission and to be precise it is better by proton induced fission than to photon induced fission, although not much significant. Sometime, the betterment is almost two orders of magnitude in the mass range 110-125 in the symmetric fission mode for incident proton energy more than 35 MeV. However, production of r-process nuclei through photon induced fission (bremsstrahlung photons from energetic electron by e-LINAC) is preferred because of better thermal management in the target design by having two targets (converter target and fission target) instead of one. But keeping a simultaneous option for proton induced fission may be a judicious choice if one really requires producing nuclei like $^{117}$Pd, $^{122}$Cd (8-9 neutron excess) or more neutron rich nuclei like $^{80}$Zn, $^{96}$Kr and $^{106}$Zr (15-16 neutron excess). 
	
	Finally, for producing neutron rich nuclei, in the higher mass range, more than A$=$160, one would need to go for proton induced reaction rather than nuclear process other than photonuclear reaction to produce them and for that high energy proton beam is required.


\noindent
\begin{thebibliography}{99}

\bibitem{Na13} H. Naik, Fr\'ed\'erick Carre, G. N. Kim, Fr\'ed\'eric Laine, Adrien Sari, S. Normand and A. Goswami, Eur. Phys. J. {\bf A 49}, 94 (2013).

\bibitem{Depp13} A. Deppman, E. Andrade-II, V. Guimarães, et al. Phys. Rev. {\bf C88}, 064609 (2013).

\bibitem{Is14} B. S. Ishkhanov and A. A. Kuznetsov, Phys. Atom. Nuclei {\bf 77}, 824 (2014). 

\bibitem{Be15} S. S. Belyshev, B. S. Ishkhanov, A. A. Kuznetsov and K. A. Stopani, Phys. Rev. {\bf C 91}, 034603 (2015). 

\bibitem{Es03} S. Essabaa et al., Nucl. Instr. and Meth. {\bf B 204}, 780 (2003). 

\bibitem{Di99} W. T. Diamond, Nucl. Instr. and Meth. {\bf A 432}, 471 (1999).

\bibitem{Bh15} Debasis Bhowmick, Debasis Atta, D. N. Basu and Alok Chakrabarti, Phys. Rev. {\bf C 91}, 044611 (2015).

\bibitem{Bh16} Debasis Bhowmick, F.A. Khan, Debasis Atta, D.N. Basu, and Alok Chakrabarti, Can. J. Phys. {\bf 94}, 243 (2016).

\bibitem{TRIUMF12} F. Ames et al., Conf. Proc. {\bf C1205201}, 64 (2012). 

\bibitem{TRIUMF11} J. Richards et al., Conf. Proc. ICALEPCS2011, Grenoble, France, 465 (2011). 

\bibitem{Dubna} Yu. Ts. Oganessian et al; Nucl. Phys. {\bf A 701}, 87 (2002).

\bibitem{Alto} M. Cheikh Mhamed et al; Nucl. Inst. Methods,  {\bf B 226}, 4092 (2008).

\bibitem{Ch13} A. Chakrabarti et al., Nucl. Instr. and Meth. {\bf B 317}, 253 (2013).

\bibitem{Jones} W. H. Jones et al., Phys. Rev. {\bf 99}, 184 (1955).

\bibitem{Stev} P. C. Stevenson et al., Phys. Rev. {\bf 111}, 886 (1958).

\bibitem{Talys} A.J. Koning, S. Hilaire, and M.C. Duijvestijn. In Proceedings of the International
Conference on Nuclear Data for Science and Technology, 22-27 April 2007, Nice, France. 2008. pp. 211-214.Arjan Koning, Stephane Hilaire and Stephane Goriely, {\bf TALYS-1.4} A nuclear reaction program, December 28, (2011).

\bibitem{Pace} O. B. Tarasov, D. Bazin, Nucl. Instr. and Meth. {\bf B 204}, 174 (2003).

\bibitem{Proko} A. V. Prokofiev et al, Nucl. Instr. and Meth. {\bf A 463}, 557 (2001).

\bibitem{Fuka91} T. Fukahori, S. Pearlstein, Proc. of advisory Group Meeting Organized by IAEA, Vienna, October 9-12, 1990, IAEA Report INDC(ND5)-245, 1991 p.93

\bibitem{Fuka97} T. Fukahori, S. Chiba, Proc. of the First Internet Symposium on Nuclear Data. April 8 - June 15, 1996, JAERI, Tokai, Ibaraki, Japan: Paper No. 09; JAERI- Conf 97-004, INDC(JPN)-178/U, p.95.

\bibitem{Letaw} J. R. Letaw, R. Silberberg, C. H. Taso, Astrophys. J. Supply. Ser. {\bf 51}, 271 (1983).

\bibitem{Baba} S. Baba et al, Nucl. Phys. {\bf A 175} 177 (1971).

\bibitem{Isa08} S. Isaev et al, Nucl. Phys. {\bf A 809}, 1 (2008).

\bibitem{Oht} T. Ohtsuki et al, Phys. Rev. {\bf C 40} 2144 (1989).

\bibitem{Rub} V. A. Rubchenya, et al, Nucl. Instr. and Meth. {\bf A 463}, 653 (2001).

\bibitem{Ch06} P. R. Choudhury and D. N. Basu Acta Phys. Pol. {\bf B 37}, 1833 (2006).

\end{thebibliography}
\end{document}